\documentclass[twocolumn,preprintnumbers,elsart]{revtex4}
\usepackage{amsmath}
\usepackage{cases}
\usepackage{mathrsfs}
\usepackage{graphicx}
\usepackage{dcolumn}
\usepackage{bm}
\usepackage[center]{subfigure}
\usepackage{color}
\usepackage{float}

\begin{document}

\title{Optical Superoscillatory Poisson-Arago Spots}
\author{Yanwen Hu$^{1}$, Shenhe Fu$^{1,2,3*}$, Zhigui Deng$^{4}$, Siqi Zhu$^{1,2,3}$, Hao Yin$^{1,2,3}$, Yongyao Li$^{5}$, Zhen Li$^{1,2,3}$, and Zhenqiang Chen$^{1,2,3}$}
\email{fushenhe@jnu.edu.cn;tzqchen@jnu.edu.cn}

\address{
$^{1}$Department of Optoelectronic Engineering, Jinan University, Guangzhou 510632, China  \\
$^{2}$Guangdong Provincial Key Laboratory of Optical Fiber Sensing and Communications, Guangzhou, 510632, China \\
$^{3}$Guangdong Provincial Engineering Research Center of Crystal and Laser Technology, Guangzhou, 510632, China \\
$^{4}$Department of Physical Electronics, Faculty of Engineering, Tel-Aviv University, Tel-Aviv, 69978, Israel \\
$^{5}$School of Physical and Optoelectronic Engineering, Foshan University, Foshan, 528000, China}

\begin{abstract}
  Optical diffraction limit has been a long-term scientific issue since Ernst Abbe first introduced the concept in 1873. It is a constraint on the smallest light spot that can be achieved. Substantial effort has been invested in the past decade to beat this limit by exploiting evanescent waves. But this method encounters serious near-field limitations. A more promising route to breaking the constraint is to explore optical superoscillation in the far field with engineered metamaterials. However, these particular structures involve with very complicated optimization-based design that requires precisely tailoring the interference of propagating waves with low spatial frequency. To overcome these limitations, here we explore a new approach based on the two-hundred-year-old discovery: Possion-Arago spots. We show for the first time that by using a single disc, constructive interference of propagating waves with high-spatial-frequency wavevectors can be realized, generating a diffraction-unlimited localized Possion-Arago spot with achievable size down to $\lambda/$20. Actually, such an element permits creation of an ultra-long nearly nondiffracting superoscillatory needle with appreciable field of view. This easy-to-fabrication element provides a promising route to overcome the diffraction limit, thus might open new avenues to exploit various applications in different fields.
\end{abstract}
\maketitle

\indent In 1818, Poisson pointed out from Fresnel's diffraction theory that there should be a bright spot of light appearing behind the center of an illuminating circular obstacle. Such a surprising prediction violates the common sense and therefore poses discredit to Fresnel's wave theory. However, Arago subsequently carried out experiment using a circular disc, and verified this prominent phenomenon \cite{Harvey1984,Pedrotti1993}. The Poisson-Arago spot shows solid evidence to the wave nature of light beam. Unfortunately, the famous discovery of the Poisson-Arago spot gradually recedes into the background as a historically effect and became almost forgotten by the scientific communities. Herein, we revisit this two-hundred-year-old phenomenon and address for the first time whether such a hotspot can beat the optical diffraction limit, generating a superoscillatory Poisson-Arago spot (SOPAS). \\
\indent Optical diffraction limit has been considered as a long-term fundamental problem \cite{Hao2013}. In 1873, Ernst Abbe first introduced the concept of optical diffraction limit \cite{Abbe1873}. It is a constraint on the smallest focal spot with size about half of the used wavelength $\lambda$. In the past decade, great effort has been devoted to breaking this limit by exploiting evanescent waves that contain high spatial frequencies with superlenses \cite{Fang2005,Davis2007,Liu2007,Zhang2008}, hyperlens \cite{Lee2007,Liu2007a,Jacob2006}, or the near-field microscopies \cite{Thomas2013,Huber2008}, e.g., the SNOM. However, these particular lenses require bulk negative index materials that, owing to the challenges of losses and unattainable fabrications, are still under investigations; while the SNOM suffers from serious near-field limitations. Recently, Berry and Popescu introduced the concept of optical superoscillation as a promising route to break the diffraction limit \cite{Berry2006}. It refers to a counterintuitive effect of a bandlimited function that contains oscillations exceeding its highest Fourier component. The most remarkable advantage of superoscillation is capable of creating arbitrary small feature size of light in the far field although it comes at the cost of generating strong sidelobe. The most common way to realize the superoscillatory spot relies on a superoscillatory lens (SOL) \cite{Huang2009,Rogers2012,Rogers2013,Luo2015,Yuan2016,Qin2017,Yuan2019,Yuan2019a}, a specially designed nanostructured mask. However, the technique of designing SOL encounters intractable problems: first, the SOL is constructed by a large number of delicate concentric belts with varying sizes in deep subwavelength, which poses great challenges in mask fabrications; second, the design of SOL relies excessively on optimization algorithms, as a result, the underlying physics of the characteristics of SOL is not clear; third, the SOL-generated superoscillatory spot is formed by interference of many propagating waves that contain low-spatial-frequency wavevectors, which gives rise to limited spot size.
\begin{figure*}[t]
\centering
\includegraphics[width=15cm]{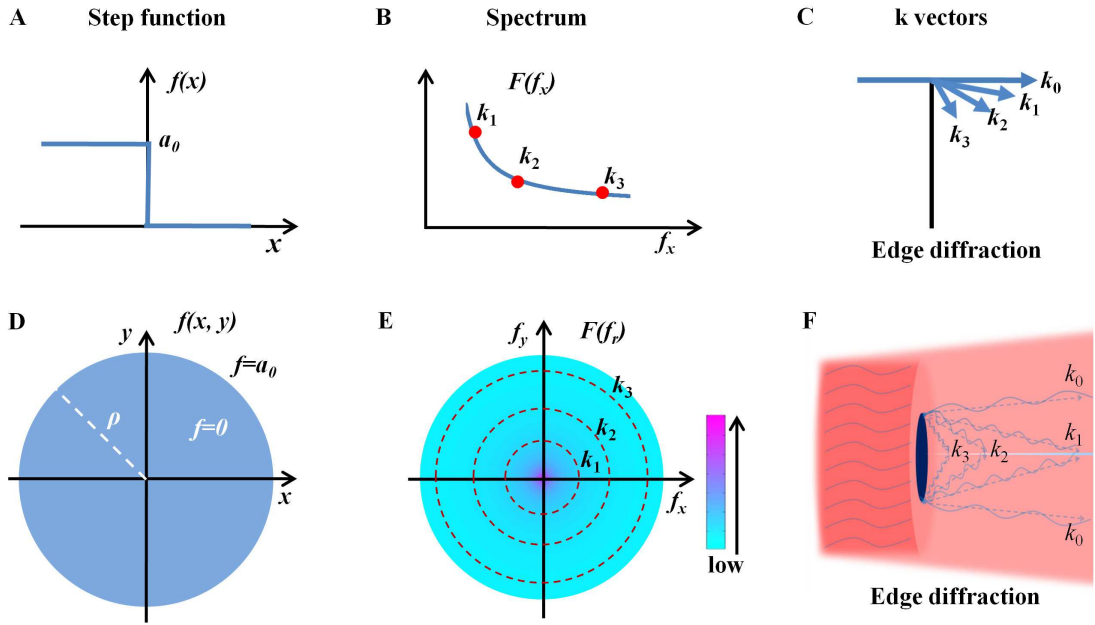}
\caption{Principle of breaking optical diffraction limit by interfering propagating waves with high spatial frequencies. (A) One-dimensional step function used to modulate amplitude of the incident waves in binary. (B) The corresponding spectrum of (A) in Fourier space, which shows significant high-spatial-frequency wavevectors, e.g., marked as $k_1$, $k_2$ and $k_3$. (C) Schematic illustration of edge diffraction of (A) showing that these wavevectors deviate from original one $k_0$. It suggests that the wave with high spatial frequency bypasses the obstacle and reach to the area on the back. In two dimensional space, a circular step function as shown in (D) possesses higher-order diffraction components having circular symmetric distribution in the reciprocal space as illustrated in (E). The high-frequency wavevectors $k_{j}$ ($j=1,2,3, ...$) located at the circular ring are in phase. Therefore they point to the same central position behind (F). As a result, it generates a constructive diffraction-unlimited spot.}
\end{figure*} \\
\indent In this article, we demonstrate both theoretically and experimentally a new route that is based on interfering propagating waves with high-spatial-frequency wavevectors to generate a superoscillatory spot in the far field. The created spot is extremely localized in space with feature size that, in principle, can be arbitrarily small. As compared to the standard SOL structures \cite{Huang2009,Rogers2012,Rogers2013,Luo2015,Yuan2016,Qin2017,Yuan2019,Yuan2019a}, this pure route relies only on a single circular disc and therefore it generates a superoscillatory Poisson-Arago spot behind the obstacle. Such an optimization-free technique has remarkable advantages: low-cost, easy-to-perform and robust to the changes of light field, e.g., the wavelength and polarization. Hence it has practical applications in various potential fields such as optical nanoscopy \cite{Rogers2012,Qin2017}, nanoparticle trapping and manipulation \cite{Singh2017} etc. Moreover, the simple element allows for studying the inaccessible limit of the superoscillatory spot size that cannot be achieved with regular SOLs.\\
\indent We start by considering the principle of using the interference of propagating waves with high-spatial-frequency wavevectors to break the optical diffraction limit. This is an extremely challenging issue. The reasons are two folds: first, the light field emitted or diffracted from an ordinary object contains few propagating waves with high spatial frequencies (the mostly high-spatial-frequency waves are the evanescent waves which decay exponentially and only exist in the near field region of the object); second, the distribution of the diffraction-induced high-spatial-frequency wavevectors in the reciprocal space, generally speaking, is irregular, which gives rise to destructive interference patterns in the far field. To overcome these issues, an object with very sharp edges and regular spatial geometry is adapted. It is expected that the sharp-edge-induced high-spatial-frequency wavevectors are significant. For example, an incident plane wave is diffracted seriously when it hits an obstacle whose object function is described by a step function $f(x)=a_0\text{step}(x)$, as illustrated in Fig. 1A. Consequently, the amplitude near the edge of the obstacle is modulated in binary. In its reciprocal space, in addition to the zero-order diffraction component, the step function corresponds to a reverse function of $f_x$ in its Fourier space, i.e., $F(f_x)=1/(i2\pi f_x)$, where $f_x$ is spatial frequency with respect to $x$ coordinate and $i$ is the imaginary unit. Figure 1B illustrates the corresponding higher-order frequency components of the step function, indicating significant higher-order wavevectors. In this scenario, these high-spatial-frequency wavevectors are able to bypass the obstacle and reach to the area on the back. Figure 1C shows three different cases of wavevectors marked as $k_1$, $k_2$ and $k_3$ deviating from the original one $k_0$, where $k_0=2\pi/\lambda$. The higher-frequency wavevector would give rise to larger deviation angle, however, the strength would become relatively weaker.
\begin{figure*}[t]
\centering
\includegraphics[width=15cm]{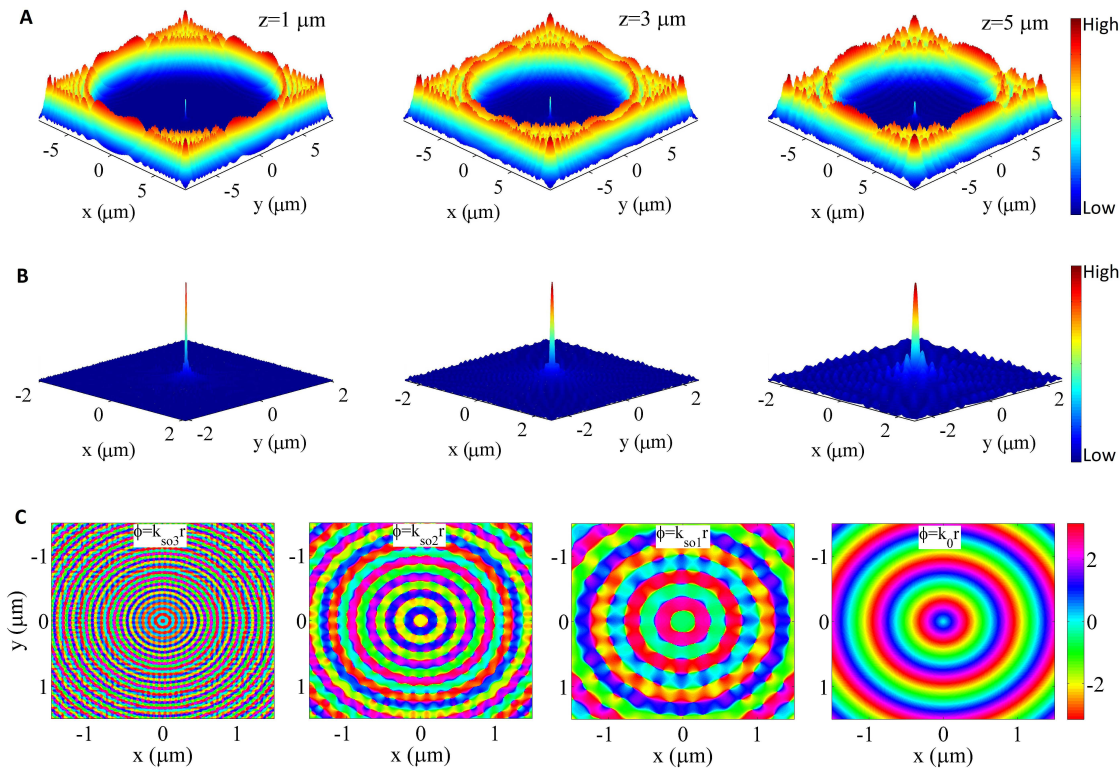}
\caption{Theoretical demonstrations of the superoscillatory Poisson-Arago spots. The highly localized spots are generated with a circular disc ($\rho=7.5$ $\mu$m), illuminated by a 633 nm laser beam. (A) The calculated intensity distributions of the diffracted light fields at propagation distance of $z=1$ $\mu$m, $z=3$ $\mu$m, and $z=5$ $\mu$m. (B) The zoom in intensity distributions of (A), in order to characterize the superoscillatory spots. The FWHMs of the spots are measured as 32 nm ($\lambda/$20), 80 nm ($\lambda/$8), and 105 nm ($\lambda/$6), respectively. (C) The spatial phase patterns of the superoscillatory fields at $z=1$ $\mu$m, $z=3$ $\mu$m, and $z=5$ $\mu$m, expressed as $\phi=k_{so3}r$, $\phi=k_{so2}r$, and $\phi=k_{so1}r$, respectively. In comparison, the spatial phase distribution given by the original wavevector: $\phi=k_{0}r$ is also presented.}
\end{figure*}\\
\indent Next, let's consider the sprungstelle of the amplitude along a circle trajectory in the x-y plane, described by a regular function: $f(r)=a_0[1-\text{circ}(r/\rho)]$, where circ($\cdot$) denotes the circle function, $r=(x^2+y^2)^{1/2}$ and $\rho$ the radius. A geometry representation of such a circular step function is presented in Fig. 1D. In the reciprocal space, such a geometry gives rise to higher-order frequency components described by a similar formula: $F(f_r)=1/(i2\pi f_r)$, where $f_r=(f_x^2+f_y^2)^{1/2}$ with $f_y$ being the spatial frequency associated with $y$ coordinate. In Fig. 1E, three different high spatial frequencies as well as their corresponding wavevectors $k_1$, $k_2$ and $k_3$ are marked as shown by the red circles. It is illustrated that, different from the one-dimensional step function, the circular step function possesses higher-order diffraction components having circular symmetric distributions in the reciprocal space. For a specific high frequency, its corresponding wavevectors located at the edge of a circular ring are in phase, having the same deviation angle and therefore pointing to the same position at the center behind the obstacle, as illustrated in Fig. 1F. As a result, it leads to constructive interference and hence produces local intensity enhancement at the center. Since the interfering propagating wave contains higher spatial frequency than that of the zero-order wavevector $k_0$, the hotspot would be highly localized in space with featue size much smaller than the diffraction limit, namely generating a superoscillatory Poisson-Arago spot.
\begin{figure*}[t]
\centering
\includegraphics[width=16cm]{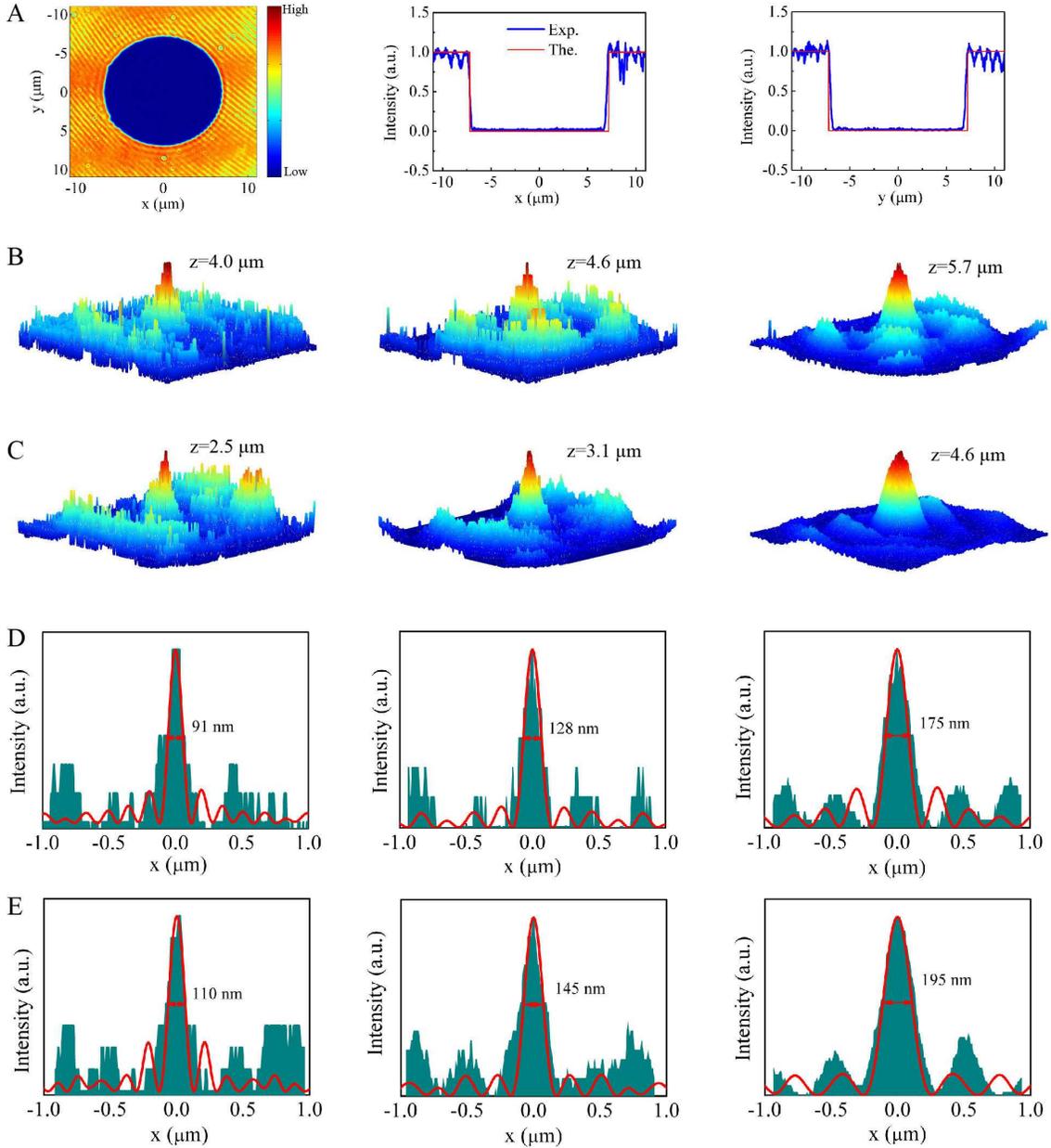}
\caption{Experimental generation of the superoscillatory Poisson-Arago spots. (A) The initially measured light field at $z=0$ as well as their corresponding intensity profiles along $x$ and $y$ coordinates indicates a binary modulation of the amplitude. (B) The experimentally measured intensity distribution of the SOPAS at distances $z=4.0$ $\mu$m, $z=4.6$ $\mu$m and $z=5.7$ $\mu$m with a metallic disc ($\rho=7.5$ $\mu$m), while (D) depicts their corresponding intensity profiles along $x$. The FWHM values of the SOPAS were measured as 91 nm, 128 nm and 175 nm, respectively. (C) and (E) present measurements of SOPAS with a smaller metallic disc ($\rho=4.5$ $\mu$m) at distances $z=2.5$ $\mu$m, $z=3.1$ $\mu$m and $z=4.6$ $\mu$m. The spot size was also measured as indicated in (E). Note that the red curves in (D) and (E) are the theoretical results based on Eq. (1).}
\end{figure*}\\
\indent We theoretically investigate the far-field diffraction patterns with such a sharp-edge geometry. The propagation dynamics of the diffracted light field $a(x,y,z)$ beyond the circular geometry that is placed at $z=0$ can be governed by the Fresnel diffraction integral
\begin{equation}
\begin{aligned}
a(x,y,z)=&\frac{a_0k_0}{2\pi z}\int\int f(x',y') \\
&\exp\left[\frac{ik_0}{2z}\left[(x-x')^2+(y-y')^2\right]\right]dx'dy',
\end{aligned}
\end{equation}
where $z>0$ represents the propagation distance. The superoscillatory performance of the geometrical element is evaluated by direct solving the equation (1). In simulations, the radius of the circular obstacle is set to $\rho=7.5$ $\mu$m, and a laser beam with $\lambda=$633 nm is utilized to illuminate the element. A 3D views of the calculated intensity distributions at the distances of $z=1$ $\mu$m, $z=3$ $\mu$m, and $z=5$ $\mu$m are presented in Fig. 2A. It is evident that in addition to dominant sidelobes caused by the $k_0$ diffractions, the diffraction patterns contain central main lobes that come from constructive interference of the in-phase higher-order wavevectors, suggesting a generation of SOPAS. In order to characterize the SOPAS, Fig. 2B depicts their corresponding central hotspots. The full widths at half maximum (FWHM) of the SOPAS are measured as 32 nm ($\lambda/$20), 80 nm ($\lambda/$8) and 105 nm ($\lambda/$6) respectively, much below the diffraction limit ($\lambda/2$). It is also shown that increasing the distance $z$ would lead to a slight increase of lateral size of the superoscillatory spot. This is because the larger diffractive distance corresponds to smaller deviation angle of the in-phase wavevectors that have relatively smaller spatial frequency. On the other hand, the superoscillatory effect of the Poisson-Arago spot can be also observed from the phase maps. Fig. 2C presents the phase distributions of the superoscillatory fields at $z=1$ $\mu$m, $z=3$ $\mu$m, and $z=5$ $\mu$m, which can be described by $\phi=k_{so3}r$, $\phi=k_{so2}r$, and $\phi=k_{so1}r$, respectively. Here $k_{so}$ is local wavevector of the superoscillatory field. It is seen that the phase of the light fields exhibits a set of circular rings with different phase values, and oscillates much more rapidly in the radial direction than the phase given by the original wavevector: $\phi=k_0r$. Since the local wavevector is defined as a gradient of phase in the $x-y$ plane \cite{Berry2006,Yuan2019,Yuan2019a}, it is easily concluded from Fig. 2C that the generated SOPAS exhibits gigantic local wavevectors $k_{so}$ with value far exceeding $k_0$. Unlike evanescent waves that only exist in the near-field surface, the SOPAS is a delicate constructive far-field interference effect of propagating waves with high spatial frequency. It provides a pure route to obtain optical spots with deeply subwavelength or even, in principle, arbitrarily small size. In simulations, due to the limited memory space, we have shown the achievable size that can be down to $\lambda/$20 (it can be even smaller if the memory space is allowed), which is inaccessible with conventional SOLs. \\
\indent To experimentally realize the SOPAS, a key issue lies in achieving binary amplitude modulation near the edge of a circular disc. Therefore, it is strictly required that the incident amplitude of the light field jumps sharply from $a_0$ to 0 at the circular edge of the element. We achieve this goal by means of fabricating a metallic disc with thickness of 60 nm on a glass substrate. A 50-nm-thick gold film was initially deposited on the substrate with magnetron sputtering. A 10-nm-thick chromium film was deposited in-between an adhesion layer. Then a mask was formed by UV lithography. The ion beam etching process was carried out to generate the final pattern. Two metallic discs of radius $\rho=7.5$ $\mu$m and $\rho=4.5$ $\mu$m were simultaneously fabricated on the same substrate to facilitate the investigations. Since the thickness of the samples is negligible as compared to the wavelength and the sample size, the spatial distribution of the light emerging from the sample can be approximately described by the Kirchhoff boundary condition: for $r<\rho$, $f=0$, while for $r>\rho$, $f=a_0$, showing a step function at the circular edge. A linearly polarized He-Ne laser beam with $\lambda=633$ nm was collimated and illuminated the sample from the substrate side. Plane wave condition is ensured by utilizing incident beam whose diameter is order of magnitude larger than the diameter of the disc. The diffracted light field beyond the samples was collected by a high-magnification objective (Nikon, 150$\times$, numerical aperture NA=0.9). Togeter with a tube lens, the light field was imaged on a CCD with pixel size of 1.4 $\mu$m. In experiment, the objective lens was mounted on an electrical-control stage with $z$-axis resolution being 20 nm, allowing for measurements of the light field at different distances.
\begin{figure}[b]
\centering
\includegraphics[width=8.5cm]{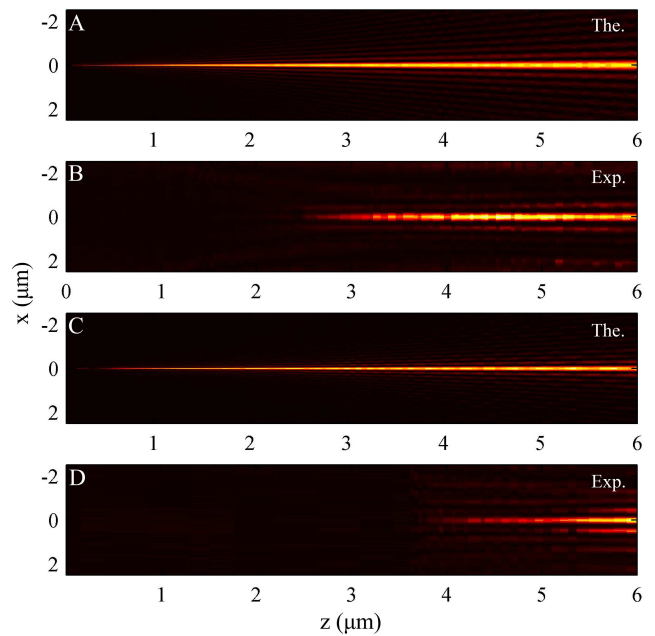}
\caption{Illustrations of extremely localized superoscillatory needles along propagation distance. The measured intensity distributions of the optical needles in the $xz$-plane under the cases of $\rho=4.5$ $\mu$m (A)(B) and $\rho=7.5$ $\mu$m (C)(D). (A)(C) Theory; (B)(D) Experiment.}
\end{figure} \\
\indent To introduce the intriguing technique to the realization of the SOPAS, we first consider the sample with $\rho=7.5$ $\mu$m. To characterize the binary modulation at the edge of the sample, we measured the intensity distribution of the light field at $z=0$, as illustrated in Fig. 3A. Their intensity profiles along the $x$ and $y$ coordinates were also presented. Apparently, the light field outside the disc is nearly invariant while approaching to the edge the field drops down sharply, showing approximately a step function at the edge. Such initial measurements validate the approximation of the Kirchhoff boundary condition. By carefully moving the stage, we obtained intensity distributions of the diffracted fields at different distances. Figure 3B plots the intensity distributions at $z=4.0$ $\mu$m, $z=4.6$ $\mu$m and $z=5.7$ $\mu$m; while Fig. 3D presents their corresponding intensity profiles along $x$ coordinate. The measured profiles of the SOPAS show good agreement with the theory depicted by red curves. The FWHM values of the central superoscillatory spots were measured as 91 nm ($\lambda/$7), 128 nm ($\lambda/$5) and 175 nm ($\lambda/$3.6) respectively. The generated SOPAS has ultrahigh spatial resolution far beyond the diffraction limit, suggesting a superoscillatory feature that the local wavevector exceeds the original one $k_0$. It is shown experimentally that increasing the distance would lead to a slight increase of the spot size, as well as the increase of energy contained in the superoscillatory main lobe. Note that the measurable minimum distance is limited by the objective numerical aperture (NA), expressed as $z_{min}=\rho(1-\text{NA}^2)^{1/2}/\text{NA}$. For $\rho=7.5$ $\mu$m, it gives $z_{min}\approx3.7$ $\mu$m. Therefore, the light field at $z<3.7$ $\mu$m cannot be collected by the objective. In the following we examine the generation of SOPAS with a smaller metallic disc: $\rho=4.5$ $\mu$m. In this case, the measurable distance is $z>z_{min}\approx2.2$ $\mu$m. Despite the change of sample size, significant superoscillatory hotspots were still observed, with measured intensity distributions depicted in Fig. 3C, as well as their intensity profiles illustrated in Fig. 3E. The FWHMs of the SOPAS at $z=2.5$ $\mu$m, $z=3.1$ $\mu$m and $z=4.6$ $\mu$m were measured as 110 nm ($\lambda/$5.8), 145 nm ($\lambda/$4.4) and 195 nm ($\lambda/$3.2) respectively.\\
\indent It is emphasized that the generation of superoscillation is inevitably accompanied by its sidelope \cite{Huang2009,Rogers2012,Singh2017,Alon2016,Dong2017,Roei2017,Kozawa2018,Lin2019}, which limits the field of view (FOV) and hence hinders its applications \cite{Rogers2012}. A tradeoff between the FOV and the size of superoscillatory spot has to be made. However, it is still challenging to design suitable SOL that generates superoscillation with appreciable feature size while maintaining large FOV. Here, the tradeoff can be easily broken via simply tuning the diameter of the disc. As indicated in Fig. 3, by increasing the size of the metallic disc, the sidelobe of the SOPAS can be easily pushed away from the center, so as to produce an applicable deeply subwavelength hotspot with large FOV. For instance, with a 9-$\mu$m-disc, it generates a SOPAS at $z=2.5$ $\mu$m with feature size of 110 nm and field of view proportional to $\rho$; whereas with a larger 15-$\mu$m-disc, it produces an even slightly smaller SOPAS at $z=4.0$ $\mu$m with a larger FOV.\\
\indent Finally, we investigate the generation of extremely localized optical needle by using the phenomenon of Poisson-Arago spot. We note that in spite of the intense research on the generation of wavepackets which are impervious to diffraction \cite{DN2007,Chong2010,Ady2013}, a strongly localized optical needle with lateral size down to deep-subwavelength scale remains a fascinating challenge \cite{Wang2008}. As illustrated in Fig. 1E, the metallic disc possesses continuous non-zero Fourier spectra, indicating a continuous high-frequency wavevectors with circular symmetric distribution in the reciprocal space. Such a unique feature permits a new approach for creating an optical superoscillatory needle in the shadow of the circular obstacle. To achieve this purpose, we carried out experiments and measured the intensity distributions along propagation distance. Figure 4 shows both the experimental and theoretical intensity distributions of the light fields in the $x-z$ plane, in the cases of $\rho=7.5$ $\mu$m and $\rho=4.5$ $\mu$m. As clearly seen, the optical needle without appreciable divergence over many Rayleigh lengths is indeed generated in free space. In theory, the needle starts at the early stage of light wave propagation. The spatial needle size decreases gradually when the Poisson-Arago spots approaching to the near-field region. Whereas in experiment, owing to the limited objective NA, the needle starts after a short distance. Even so, the experimentally measured needles match well with the theory for both cases of samples. Slight difference between the experiment and theory is explained by approximate fulfillment of the binary modulation at the edge of the sample, as can be judged from Fig. 3A. The nearly non-diffracting phenomenon of the superoscillatory needle suggests that the lateral size of the SOPAS increases very slowly with distance. This does not contradict to Berry's prediction that a superoscillatory wavepacket disappears eventually during evolution \cite{Berry2006}. Despite the slight increase of the SOPAS size, the superoscillatory feature can be maintained up to a distance $z_{so}$ that is many Rayleigh lengths. For a small disc of $\rho=4.5$ $\mu$m, it can be calculated as $z_{so}=6$ $\mu$m where the superoscillatory spot size is up to the diffraction limit $\lambda/2$; while for $\rho=7.5$ $\mu$m, it surprisingly increases to a significant value of $z_{so}=11$ $\mu$m, a distance more than 55$z_R$ where $z_R=\pi$FWHM$^2/\lambda$ is the Rayleigh range. \\
\indent It is worth mentioning that owing to the wave nature of diffraction, any spatially localized wavepacket tends to broaden, and therefore it is extremely challenging to achieve an optical spot as well as a nondiverging optical needle in free space that can break optical diffraction limit. The phenomenon of the Poisson-Arago spots, however, provides a pure route to break the diffraction limit. We realized this intriguing superoscillatory Poisson-Arago spot with an extremely thin metallic disc which shows excellent binary modulation of the plane wave amplitude and thus produces significant high-spatial-frequency wavevectors. Our demonstration provides a complete and consistent picture for achieving such a highly localized spot. These are inaccessible with optimization-based SOLs that involve with complicated structural designs and are sensitive to the changes of light parameters such as wavelength and polarization \cite{Huang2009,Rogers2012,Rogers2013,Luo2015,Yuan2016,Qin2017,Yuan2019,Yuan2019a}. In addition, we have shown that the single thin disc can create an ultra-long nearly nondiverging optical needle with controllable field of view, which ensures high tolerance in axial position regarding to a scanning mode in superresolution microscopy \cite{Rogers2012,Qin2017}. The easy-to-fabrication and -integration disc opens new avenues to explore applications in various potential fields such as superresolution microscopic imaging \cite{Rogers2012,Qin2017,Jia2014}, light field manipulations \cite{Wang2008}, nanoparticle manipulations \cite{Singh2017}, etc.
\newline\newline
\noindent\textbf{ Acknowledgements} \\
\noindent \textbf{Funding:} This work was supported by the National Key Research and Development Program of China (2017YFB1104500), the National Natural Science Foundation of China (11704155, 11974146, 61935010), the Natural Science Foundation of Guangdong Province (2017B030306009), The Science and Technology Planning Project of Guangdong Province (2018B010114002), and the Pearl River talent project (2017GC010280). \textbf{Author contributions:} S. F. conceived the idea of this work. Y.H. and S. F. performed the measurements and wrote the paper. All the authors contributed to the design of the experiments and to the interpretation of the results. Z. C. supervised all the work. \textbf{Completing interest:} The authors declare no competing interests. \textbf{Data and materials availability:} All data are available in the manuscript and the supplementary materials.

\end{document}